\documentclass[fleqn,usenatbib]{mnras}
\usepackage[utf8]{inputenc}
\usepackage[catalan, english]{babel}
\usepackage{graphicx}
\graphicspath{{./figures/}}
\usepackage{natbib}
\usepackage{appendix}
\usepackage{tabularx}
\usepackage{color}
\usepackage{booktabs}
\usepackage{threeparttable}
\usepackage{longtable}
\usepackage{multicol}
\usepackage{multirow}
\usepackage[normalem]{ulem}
\usepackage{enumerate}
\usepackage{amsmath}
\usepackage{amssymb}
\usepackage{setspace} 
\usepackage{hyperref}
\usepackage[english]{cleveref}
\usepackage{float}

\definecolor{halfgray}{gray}{0.55}
\definecolor{naviblue}{RGB}{0,0,102}
\definecolor{webbrown}{rgb}{.6,0,0}
\definecolor{RoyalBlue}{cmyk}{1, 0.50, 0, 0}
\definecolor{webgreen}{rgb}{0,.5,0}
\definecolor{Maroon}{cmyk}{0, 0.87, 0.68, 0.32}
\definecolor{Black}{cmyk}{0, 0, 0, 0}
\definecolor{myorange}{RGB}{239, 186,67}

\hypersetup{%
    colorlinks=false,linktoc=all, pdfstartpage=1, pdfstartview=FitV,%
    breaklinks=true, pdfpagemode=UseNone, pageanchor=true, pdfpagemode=UseOutlines,%
    plainpages=false, bookmarksnumbered, bookmarksopen=true, bookmarksopenlevel=2,%
    hypertexnames=true, pdfhighlight=/O,
    urlcolor=webbrown, linkcolor=RoyalBlue, citecolor=webgreen,
}

\crefname{equation}{equation}{equations}
\crefname{figure}{figure}{figures}
\crefname{appsec}{appendix}{appendices}
\creflabelformat{equation}{#2#1#3}

\newcommand{\lyb}{Ly$\beta$}
\newcommand{\lya}{Ly$\alpha$}
\newcommand{\siiv}{Si\thinspace IV}
\newcommand{\civ}{C\thinspace IV}

\newcommand{\ciii}{C\thinspace III}
\newcommand{\neiv}{Ne\thinspace IV}

\newcommand{\angs}{\textup{\AA}}

\begin{document}

\title{Spectroscopic QUasar Extractor and redshift (z) Estimator SQUEzE II: Universality of the results}
\date{}
\author[Ignasi P\'erez-R\`afols et al.]
  {Ignasi ~P\'erez-R\`afols,$^{1,2}$\thanks{email: ignasi.perez@lam.fr}, Matthew M. Pieri$^{1}$\\
$^{1}$Aix Marseille Univ, CNRS, CNES, LAM, Marseille, France\\
$^{2}$Sorbonne Université, Université Paris Diderot, CNRS/IN2P3, Laboratoire de Physique Nucléaire et de Hautes Energies, LPNHE,\\
4 Place Jussieu, F-75252 Paris, France\\
}

\maketitle

\begin{abstract}
  In this paper we study the universality of the results of SQUEzE, a software package to classify quasar spectra and estimate their redshifts. The code is presented in \cite{Perez-Rafols+2019}. We test the results against changes on signal-to-noise, spectral resolution, wavelength coverage, and quasar brightness. We find that SQUEzE levels of performance (quantified with purity and completeness) are stable to spectra that have a noise dispersion 4 times that of our standard test sample, BOSS. We also find that the performance remains unchanged if pixels of width 25$\angs$ are considered, and decreases by $\sim2\%$ for pixels of width 100$\angs$.  We see no effect when analysing subsets of different quasar brightness, and we establish that the blue part (up to 7000\AA) of the spectra is sufficient
for the classification. Finally, we compare our suite of tests with samples of spectra expected from WEAVE-QSO and DESI, and narrow-band imaging from J-PAS. We conclude that SQUEzE will perform similarly when confronted with the demands of these future surveys as when applied to current BOSS (and eBOSS) data.

\end{abstract}


\section{Introduction}\label{sec:intro}

Current spectroscopic surveys are generating hundreds of thousands of spectra of objects targeted as quasars. These spectra need to be inspected to determine whether the observed object is indeed a quasar, and to estimate their redshift when that is the case. One of such surveys is the Baryon Oscillation Spectroscopic Survey \citep[BOSS;][part of the Sloan Digital Sky Survey - III survey \citealt{Eisenstein+2011}]{Dawson+2013}, with 546,856 spectra targeted as quasars that were visually inspected \citep{Paris+2017}. In light of next generation of surveys such as WEAVE-QSO (\citealt{Pieri+2016}; as part of WEAVE, \citealt{Dalton+2016}), DESI \citep{DESI2016}, Euclid \citep{Laureijs+2010} and J-PAS \footnote{J-PAS is an imaging survey survey with 56 narrow band-filters as can be treated as effectiveluy providing `pseudo-spectra'} \citep{Benitez+2014}, it has become clear that these surveys will provide potential quasar (pseudo-)spectra in numbers too large for a survey-wide visual inspection to be viable.

In \cite{Perez-Rafols+2019}, hereafter Paper I, we presented a code, SQUEzE\footnote{Publicly available at \url{https://github.com/iprafols/SQUEzE}}, to automatically inspect all these spectra. SQUEzE works by measuring the presence and relative strength of potential quasar emission lines in coarse bands of spectrum using a series of metrics. The performance of the algorithm is tested using BOSS data and yields a purity of $97.40\pm0.47\%$ ($99.59\pm0.06\%$ for quasars with $z \geq 2.1$) and a completeness of $97.46\pm0.33\%$ ($98.81\pm0.13\%$ for quasars with $z \geq 2.1$) when a confidence threshold of $p_{\rm min}=0.32$ is used. Here (and throughout this paper), purity is defined as the number of quasars that are correctly classified divided by the total number of objects called quasars, and completeness as the number of quasars correctly classified divided by the total number of quasars that were fed to SQUEzE. Note that for a quasar to be correctly classified we require its redshift to be within 0.15 of the `true' redshift (defined currently as that which is given by visual inspection). Other codes, such as QuasarNet \citep{Busca+2018} or RedRock (an unpublished DESI code that develops the methods used by the BOSS pipeline; see \citealt{Bolton+2012}), 
appear to have roughly similar performance to SQUEzE
(a detailed comparison between the performance of SQUEzE and these codes will be given in follow-up papers in the series), but  access the entire spectra. The critical decision making algorithms of SQUEzE, on the other hand, use only high-level metrics. The difference is noteworthy because all current training and performance tests have made based on BOSS data (the only large enough visually inspected datatset available), while these codes will be applied to {\it other} surveys with different resolution, noise properties, pipeline reduction software. The minimal direct access of SQUEzE to the spectra makes it more resistant to these potential peculiarities. 

SQUEzE sets only minimal requirements on the data: a spectral resolution sufficient to marginally resolve quasar emission lines and a wavelength range wide enough to access a small number of detectable emission lines simultaneously at the redshifts of interest.
This versatility is a key element of SQUEzE, as the main goal of these classifiers is to operate on data from new surveys, rather than from the already complete BOSS survey. We test the limits of this versatility with low signal-to-noise (increasing the noise dispersion by a factor of up to 4), low resolution (down to pixels of 100$\angs$ in width), and changes in quasar properties with brightness.

Our tests are not arbitrarily chosen. They refer in broad terms to the next generation surveys stated above. All these quasar samples are dominated by their faint limits and  identification is limited by the success at these faint limits. Both WEAVE-QSO and DESI will acquire spectra of $z>2.1$ quasars with signal-to-noise $\gtrsim 0.4$ at full depth, and this is the approximate minimum signal-to-noise of BOSS quasar spectra. A further challenge is presented by DESI since it must identify and perform a redshift estimate on essentially all quasars targeted with an exposure time 4 times shorter, and this corresponds to our 4 times noise test limit. Finally J-PAS will acquire narrow band imaging data of quasars with filters centres separated by 100$\angs$ broadly equivalent 100$\angs$ binning, and will reach depth necessary to identify WEAVE-QSO targets ($r<23.2$), which corresponds to our 4 times noise test limit. Finally some target selection from J-PAS data will be acquired at times by only partial wavelength coverage, addressed by our other tests.

We therefore estimate the performance of the algorithm when applied to datasets with different properties, and determine the conditions where the trained algorithm is no longer applicable. Using only high-level metrics, SQUEzE is resistant to the changes mentioned above. The object of this paper is to quantify the extent to which these changes affect the performance of SQUEzE. We will first address how performance is affected by changes related to the usage of different instrumentation: changes in the signal-to-noise, changes in spectral binning (and effectively resolution), and changes in wavelength coverage. We also explore the impact on performance when studying samples of fainter quasars. Finally we explore the challenge of acquiring faint quasar samples from J-PAS data by combining both increased noise and decreased resolution.

As in Paper~I, in this paper we assume that the visual inspection is always correct. Based on a preliminary visual re-inspection of a subset of the apparent failures, we believe that the visual inspection catalogue suffers from low levels ($\sim 1\%$) of impurity and incompleteness. Therefore a more detailed study of the validity of this statement will be addressed in a follow-up paper on the series. The datasets used here are the same used in Paper~I and are described in Section 3 of Paper~I, but we summarize its properties here. We use 8 independent pairs of training and validation samples. Each of the samples consists of 64 plates of BOSS ($\sim6,800$ quasars and $\sim11,520$ contaminants), and all 16 samples are independent from each other. The spectra in these samples are modified for the different tests as explained in the corresponding sections.

This paper is organised as follows. We start by giving an overview of SQUEzE in 
Section~\ref{sec:review}. Then we move to testing changes in signal-to-noise in Section~\ref{sec:noise}, changes in spectral binning in Section~\ref{sec:binning} and changes in wavelength coverage in Section~\ref{sec:wave_coverage}. Tests on the effect in changes on quasar properties are performed in Section~\ref{sec:brightlow}. Finally, we discuss the performance of SQUEzE on upcoming surveys in Section~\ref{sec:new_surveys} and summarize our results in Section~\ref{sec:discussion}.

\section{SQUEzE overview}\label{sec:review}
In this section we review the SQUEzE algorithm. The purpose of this section is not to provide a detailed description of the code (this is done in Paper~I), but to summarize its main features, and introduce the key concepts used throughout the paper.

In order to reproduce the essential elements of visual inspection, SQUEzE follows a three-step procedure. In the first step, the peak finder locates all significant peaks on a given spectrum. Each of these peaks is 
assigned a trial emission line classification (see Table~\ref{ta:lines}). Every trial classification for every peak, implies a trial redshift, $z_{\rm try}$, and an accompanying set of  high-level metrics are computed. We refer the reader to Paper~I for a detailed definition of these metrics. Note that that spectra without any significant peaks are rejected at this stage and never get processed by the random forest classifier.

\begin{table}
	\centering
	\begin{tabular}{cc}
		\toprule
		\midrule
		line & wavelength [\AA\ ] \\
		\midrule
		\lyb{} & 1033.03 \\
		\lya{} & 1215.67 \\
		\siiv{} & 1396.76 \\
		\civ{} & 1549.06 \\
		\ciii{} & 1908.73 \\
		\neiv{} & 2423.83 \\
		\bottomrule
	\end{tabular}
	\caption{Line's name and nominal wavelength  for the selected lines. All wavelengths are given in $\angs$ at the restframe.}
	\label{ta:lines}
\end{table}

After this first step the list of trial redshifts and accompanying metrics are fed into a random forest classifier {\bf that} will determine whether the trial quasar and redshift classification does indeed correspond to a quasar at that redshift, and assigns a confidence level for this classification. Note that Lyman $\alpha$ forest quasar candidates (i.e. those with $z_{\rm try}\geq2.1$) are classified separately from lower redshift candidates.  We treat a classification as correct if the redshift is within 0.15 of the true redshift, $z_{\rm true}$ (given by the visual inspection). 

SQUEzE third and final step consists of choosing for each spectrum, the preferred classification and constructing the final catalogue. To construct the final catalogue, a confidence threshold, $p_{\rm min}$, is set. Objects with lower confidence values will not be included in the catalogue. Naturally, purity and completeness are affected by the choice of $p_{\rm min}$. Higher values of $p_{\rm min}$ correspond to samples that are purer and less complete. We follow the standard we set in Paper I and focus on the case where the confidence value gives approximately equal purity and completeness for the entire quasar sample (hence the $z>2.1$ subset is not deliberately balanced here). Other choices are possible based on the needs of the user.

At this point we want to introduce the concept of line confusion (this will be relevant in Section~\ref{sec:binning}). This phenomenon occurs when the wrong emission line is assigned to a peak. In these cases, the trial redshift error will be too large, and we expect it not to be classified as a quasar. However, because the wavelength ratio  of lines is sometimes similar, some trial redshifts originating from assigning the wrong emission line are still classified as quasars.

\section{Performance vs signal-to-noise}\label{sec:noise}
As stated above, we want to estimate the performance of SQUEzE when applied to datasets of different properties. The first property we change is the signal-to-noise. We modify the signal-to-noise in BOSS data by adding Gaussian noise to the spectra. The noise is added at the pixel level as
\begin{align}
    \label{eq:add_noise}
    & f'_{i} = f_{i} + \left(N_{\rm noise}-1\right)\sigma_{i}G(0, 1) ~,\\
    & \sigma'_{i} = \sigma_{i}\sqrt{N_{\rm noise}} ~,
\end{align}
where $f_{i}$ and $\sigma_{i}$ are the flux and standard deviation 
in a given pixel, $f'_{i}$ and $\sigma'_{i}$ are the modified fluxes and standard deviations, $G(0, 1)$ is a random number drawn from a Gaussian distribution with mean 0 and standard deviation 1.

\begin{figure}
    \centering
    \includegraphics[width=0.47\textwidth]{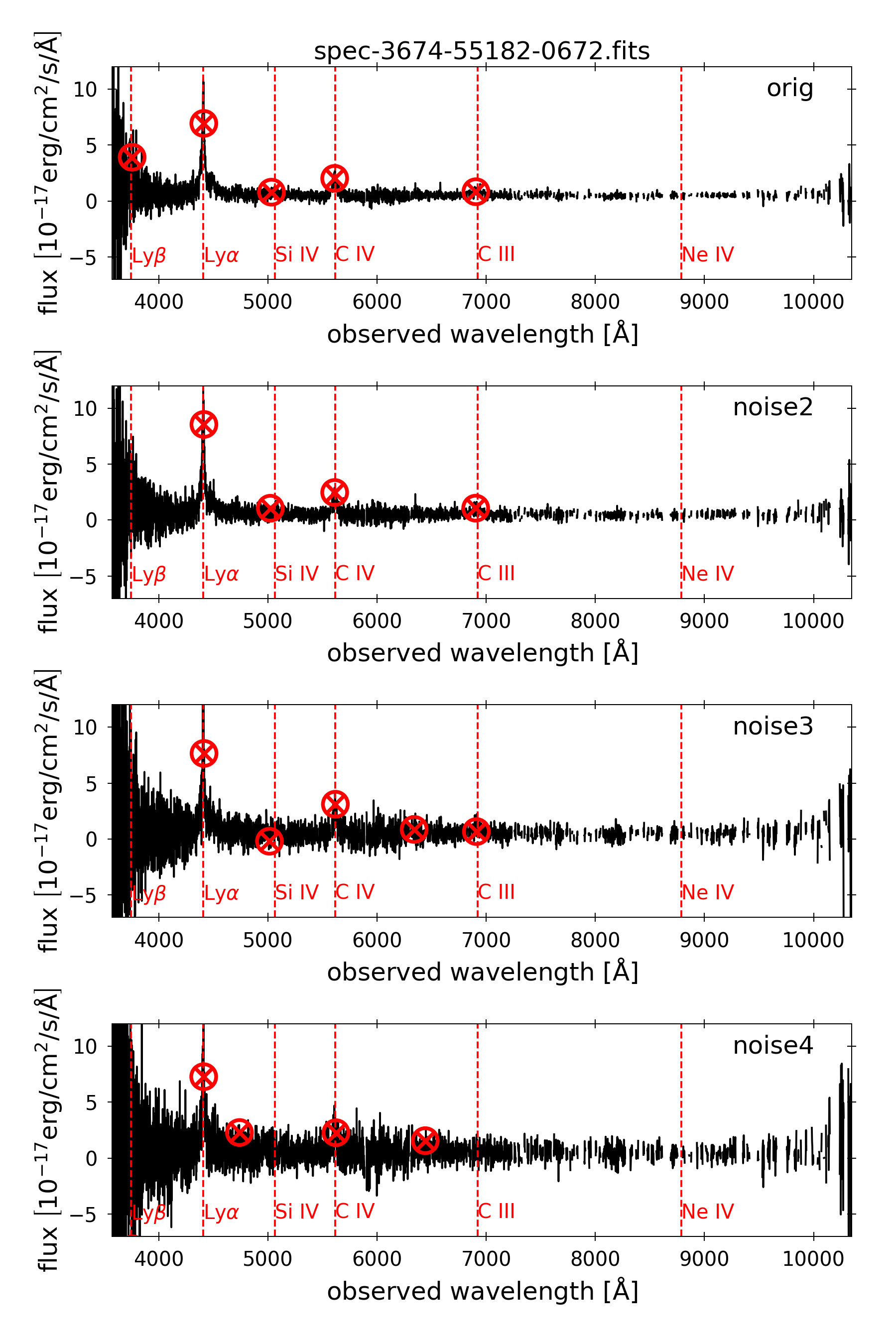}
    \caption{Example of the performance of the peak finder with spectra with different signal-to-noise. The noise is added to the spectrum following the prescription in equation~\ref{eq:add_noise}. The cases noise2, noise3, and noise4 correspond to increasing the noise by a factor of 2, 3, and 4 respectively. Dashed lines show the emission lines in Table~\ref{ta:lines} redshifted to the observed wavelength using the visual inspection reshift of the quasar. Gaps are present due missing or unreliable data (for example due to sky lines).}
    \label{fig:peak_finder_vs_noise}
\end{figure}

The case $N_{\rm noise}=1$ corresponds to the original data, and we explore the cases $N_{\rm noise}=2, 3, 4$, corresponding to having 2, 3, and 4 times the original noise. We name these cases  noise2, noise3, and noise4. Figure~\ref{fig:peak_finder_vs_noise} illustrates the noise addition in a randomly selected spectrum and the behaviour of the peak finder in each of the cases. We can see that in the original spectrum we successfully identify several emission lines: \lyb{}, \lya{}, \siiv{}, \civ{}, and \ciii{} (see the restframe wavelength for these lines in Table~\ref{ta:lines}). When we increase the noise we maintain the detection of \lya{} and \civ{}, whereas we lose the weaker \lyb{} and \siiv{} lines.  As noise increases, the peak finder fails to detect the less prominent peaks and begins to select noise spikes.  However, since spectra are smoothed before the peak search is performed, some real peaks are still found even as the noise is increased.

\begin{figure*}
    \centering
    \includegraphics[width=\textwidth]{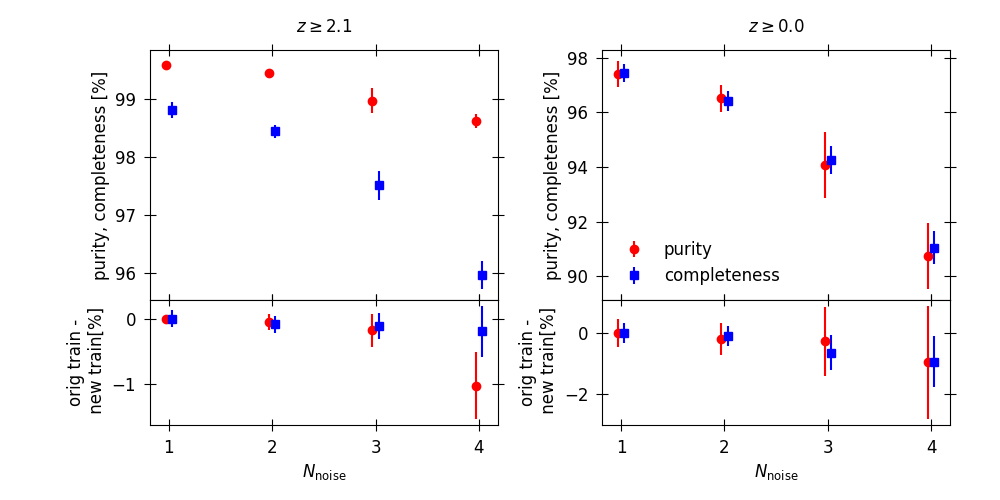}
    \caption{Top panels: Purity and completeness as a function of the noise level (see equation \ref{eq:add_noise}). For each sample a new model is retrained on the modified data, and we select a $p_{\rm min}$ such that purity and completeness are similar for the entire sample and the right panels show the performance for these samples. The left panels show the performance limiting ourselves to quasars with $z\geq2.1$. Points are horizontally shifted to avoid overlap. Bottom panels: difference in the performance between using the original or the retrained models. Negative (positive) values indicate that the retrained (original) models are better. $N=1$ correspond to the original data in Paper I.}
    \label{fig:noise_summary}
\end{figure*}

We now explore the performance of SQUEzE on these noise-augmented spectra. We ran SQUEzE twice, first retraining the models on the modified data, and then using the models obtained in the original training. Figure~\ref{fig:noise_summary} shows the results of this test taking values of $p_{\rm min}$ such that purity is approximately equal to completeness for the entire sample (also the standard choice in Paper I). 
In the top panel, using the retrained models, we see that the performance decreases by $\sim1\%$ for case noise2, by $\sim3-4\%$ for case noise3, and by $\sim7\%$ for case noise4. While the decrease is expected, we note that SQUEzE is obtaining purity and completeness of around 91\% even when the noise 
dispersion is 4 times that of the original sample. 
For \lya{} quasars (with $z \geq 2.1$) the results are even more stable: purity decreases by only $\sim1\%$ for case noise4, while completeness decreases by $\sim3\%$.
In the bottom panel we show the difference in performance when using the original or the retrained models, where we see that there is essentially no need to retrain if small changes on the noise levels are expected. If the sample is significantly noisier, as is the case for case noise4, then retraining the models minimizes losses. 

\begin{figure*}
    \centering
    \includegraphics[width=\textwidth]{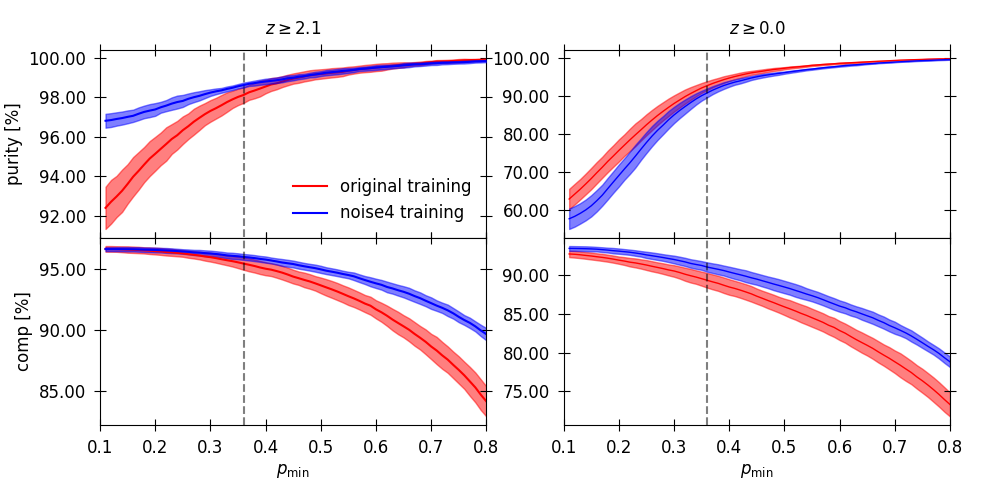}
    \caption{The dependence of the noise4 results using the original or the retrained models for varying probability thresholds. The plot shows the purity (top panels) and the completeness (bottom panels) measured on the test sample with the noise variance increased by a factor of 4.
    Red and blue lines show the results when the classifier is trained with the original data and with the noise-augmented data, respectively. Left panels shows the results for $z_{\rm try} \geq 2.1$, whilst right panels show the results at all redshifts. Dashed vertical lines show the $p_{\rm min}$ used in Figure~\ref{fig:noise_summary}.}
    \label{fig:noise2}
\end{figure*}

The above assessment of additional noise refers only to the choice of confidence threshold where the purity and completeness are approximately equal. We may broaden the exploration to the different choices of confidence threshold.  Figure~\ref{fig:noise2} shows this focusing on the difference between unmodified noise and our largest noise test, noise4. Again we see that additional noise has a weak effect even when we retain the training model generated from the unmodified data, and that weak effect is made even smaller by retraining.

\section{Performance vs binning/resolution}\label{sec:binning}
\begin{figure*}
    \centering
    \includegraphics[width=0.47\textwidth]{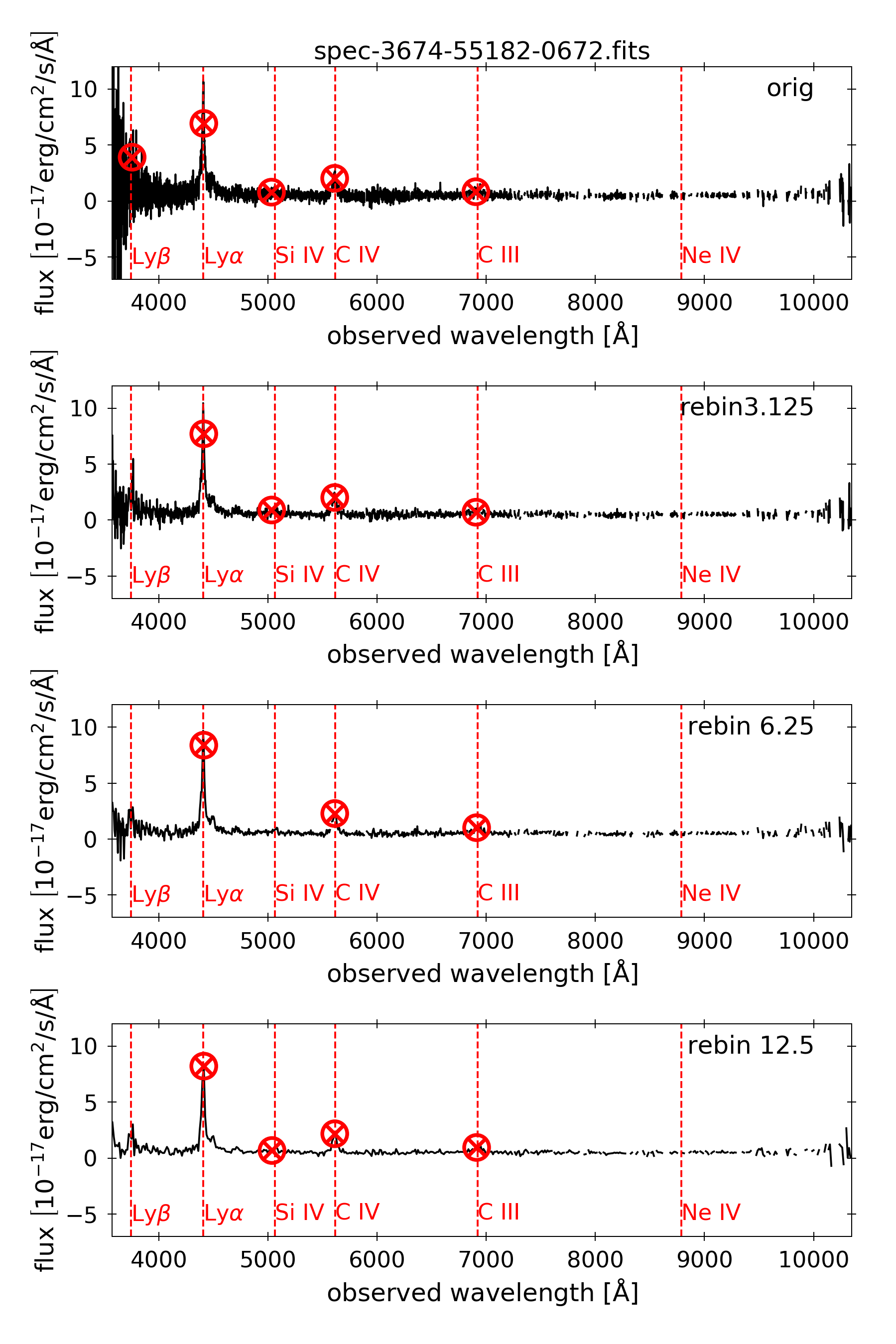}
    \includegraphics[width=0.47\textwidth]{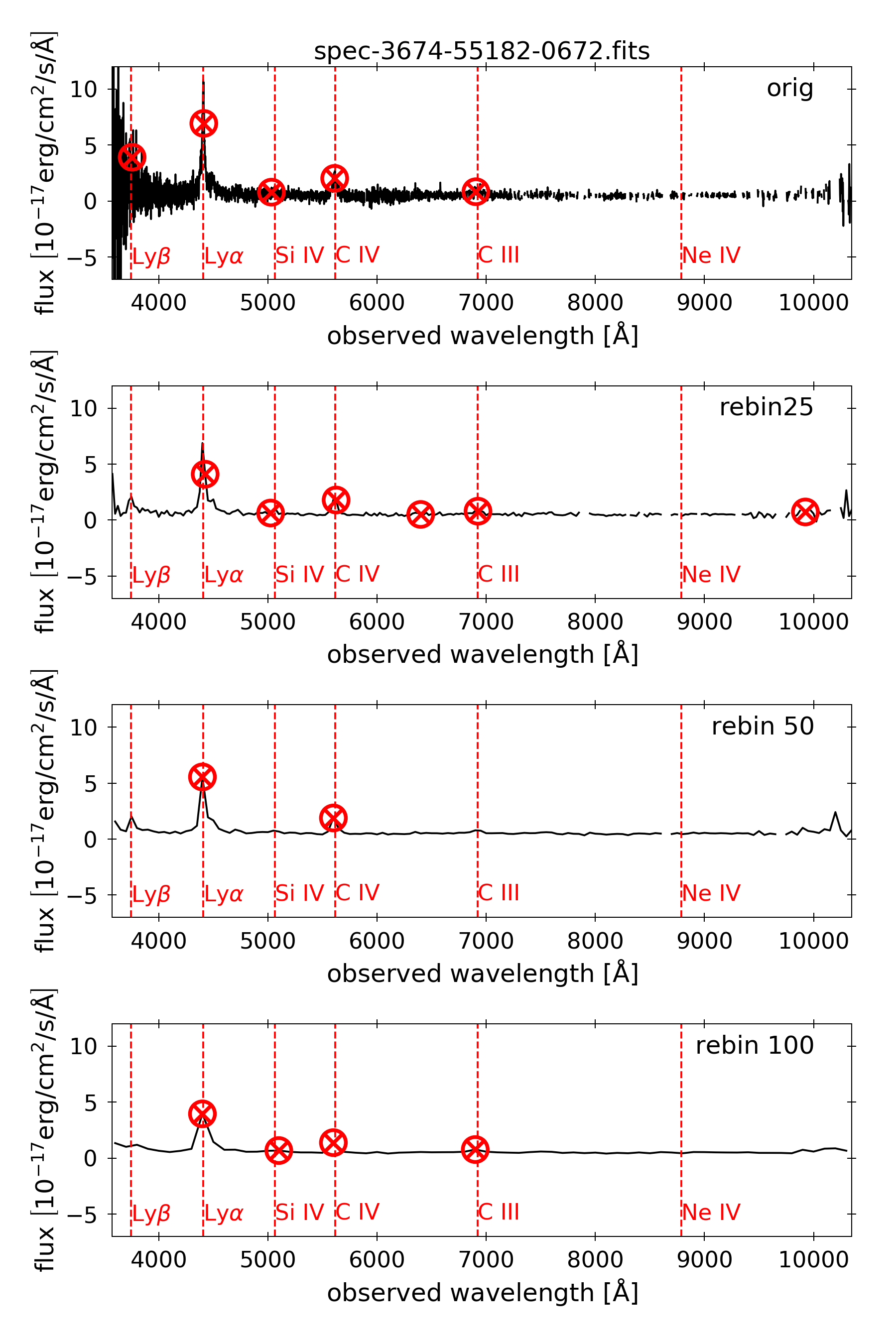}
    \caption{Example of the performance of the peak finder with spectra with different resolution.}
    \label{fig:peak_finder_vs_rebin}
\end{figure*}

Now that we have seen that SQUEzE is insensitive to changes in the signal-to-noise, we shift our attention to changes in the resolution. Now we modify the original data by rebinning the data into wider bins. The rebinned flux is computed by averaging fluxes into wider bins, and the rebinned error estimates are computed by standard error propagation. The new bins are created so that there is a bin centered at 4,000$\angs$ and have widths 3.125, 6.25, 12.5, 25, 50, and 100$\angs$. We name these cases rebin3.125, rebin6.25, rebin12.5, rebin25, rebin50, and rebin100 respecrively. For reference, the size of the original pixels is $\sim1\angs$. Figure~\ref{fig:peak_finder_vs_rebin} illustrates this rebinning and the behaviour of the peak finder for the same example spectrum as used in  Figure~\ref{fig:peak_finder_vs_noise}. Again, we can see that in the original spectrum 5 emission line peaks are successfully identified. As the bins become broader, this example shows that the peak finder no longer retains the \lyb{} line in the rebin100 test as the peak height is diminished through the dilution of local averaging. Also we see that there is some instability in the peak detection for peak 25, 50 and 100 \AA{} for our example spectrum, but the emission lines in question were not needed for classification. Note that the scale of smoothing performed by the peak finder before looking for peaks is decreased as the pixel size increases (see Table~\ref{ta:rebin}).

\begin{table}
    \centering
    \begin{tabular}{ccc}
    \toprule
    Case study & Pixel size $\left[\angs\right]$ & Smoothing [num bins] \\
    \midrule
    original & $\sim 1$ & 70 \\
    rebin3.125 & 3.125 & 22 \\
    rebin6.25 & 6.25 & 11 \\
    rebin12.5 & 12.5 & 6 \\
    rebin25 & 25 & 3 \\
    rebin50 & 50 & 1 \\
    rebin100 & 100 & 0 (no smoothing) \\
    \bottomrule
    \end{tabular}
    \caption{Bin size and kernel of smoothing applied by the peak finder before looking for peaks for the different case studies involving changes in resolution. New bins are made so that there is a bin centered at 4,000$\angs$.}
    \label{ta:rebin}
\end{table}

\begin{figure*}
    \centering
    \includegraphics[width=\textwidth]{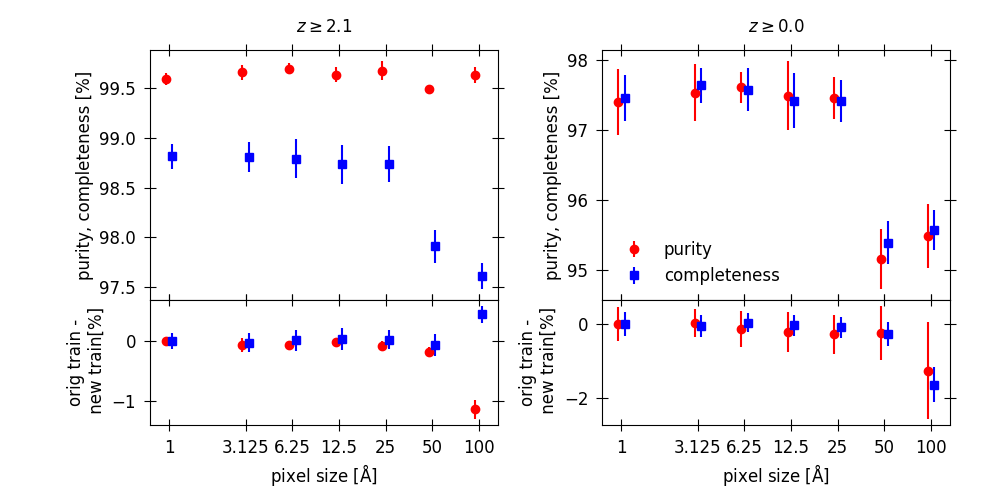}
    \caption{Similar to Figure~\ref{fig:noise_summary} but as a function of the resolution. Pixel size of 1$\angs$ correspond to the original data. }
    \label{fig:rebin_summary}
\end{figure*}

Once we rebinned the spectra we ran SQUEzE on them to assess its performance. As before we ran SQUEzE twice, both retraining the model on the modified data, and using the original training. We show the results of this test in Figure~\ref{fig:rebin_summary}, where again we choose values of $p_{\rm min}$ such that purity and completeness are similar. We see that there is no change in the performance for bins of sizes up to 25$\angs$. Up to bins of this scale, retraining the model does not improve the performance. The performance starts decreasing for bin size of 50$\angs$, but is only $\sim2\%$ lower even for bin size of 100$\angs$. Using the original training has no effect for the case rebin50 but decreases the performance of SQUEzE by an extra $\sim 2\%$ for the case rebin100. For \lya{} quasars the decrease in the performance is less prominent: purity is not modified and completeness drops by $\lesssim1.5\%$. We conclude that SQUEzE is resistant to changes in the resolution, provided that the pixel size is smaller than 25$\angs$. When binning further to 100\AA, retraining based on equivalent data helps alleviate the decrease in performance.

\begin{figure*}
    \centering
    \includegraphics[width=\textwidth]{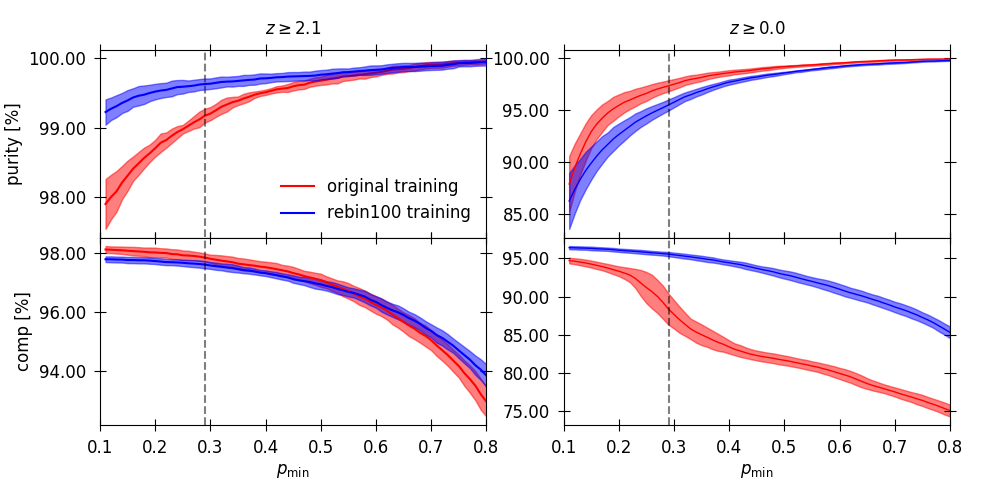}
    \caption{The dependence of the rebin100 results using the original or the retrained models for varying probability thresholds. The plot shows the purity (top panels) and the completeness (bottom panels) measured on the test sample with pixels rebinned to $100\angs$-wide pixels.
    Red and blue lines show the results when the classifier is trained with the original data and with the rebinned data, respectively. Left panels shows the results for $z_{\rm try} \geq 2.1$, whilst right panels show the results at all redshifts. Dashed vertical lines show the $p_{\rm min}$ used in Figure~\ref{fig:rebin_summary}.}
    \label{fig:rebin100}
\end{figure*}

We now analyze in more detail the case rebin100. As already seen in Figure~\ref{fig:rebin_summary} the performance of SQUEzE is decreased in this case study, and the decrease is significantly worse if the original training is used when the confidence threshold is set such that purity and completeness are similar. In Figure~\ref{fig:rebin100} we explore whether this statement holds for varying confidence threshold $p_{\rm min}$. We see that retraining improves the completeness for the all redshifts quasar sample, particularly for high values of $p_{\rm min}$. We note that purity significantly increases when using the retrained model (with respect to the purity obtained using the original model) for $z \geq 2.1$ at low values of $p_{\rm min}$.

\begin{figure}
    \centering
    \includegraphics[width=0.47\textwidth]{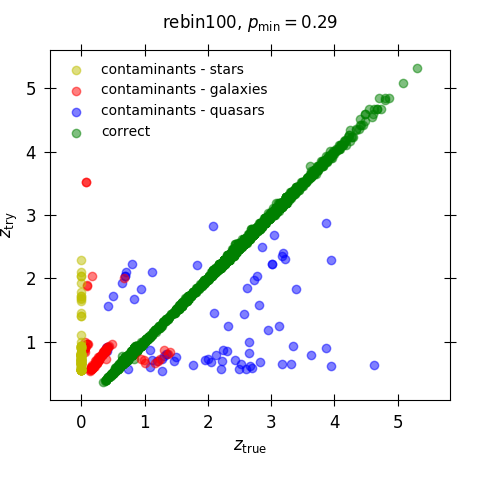}
    \caption{Line confusion plot for the case study rebin100, where the model has been retrained. For all the objects in the catalogue, $z_{\rm try}$ against $z_{\rm true}$. In this plot, green circles correspond to correct classifications, yellow circles to stellar contaminants, red circles to galactic contaminants, and blue circles to quasar contaminants (i.e. actual quasars classified as quasars but not meeting our redshift requirements). Note that this plot singles out the false positives.}
    \label{fig:line_confusion_rebin100}
\end{figure}

A possible (naive) explanation for the decrease in the performance of SQUEzE in the case rebin100 would be that the rebinning is moving the center of the emission peaks, and therefore the estimated quasar redshift. In this scenario, quasar redshift errors
would be larger than the tolerance redshift, $\Delta z_{r}=0.15$, discarding some quasars that are correctly identified. The validity of this scenario can be easily seen in a so called line confusion plot. A line confusion plot shows $z_{\rm try}$ as a function of $z_{\rm true}$ and highlights the presence of line confusion, seen as straight lines trends in the failure cases far from the $z_{\rm try} = z_{\rm true}$ line and with slope different from unity.
If the naive explanation were correct, we would see quasar contaminants (i.e. actual quasars classified as quasars but not meeting our redshift requirements) very close to the $z_{\rm try} = z_{\rm true}$ line in the line confusion plot (see Figure~\ref{fig:line_confusion_rebin100}). Since we do not see such an effect, we can rule this out. Thus, it appears that at this level of binning the calculation of the metrics themselves becomes somewhat compromised since the bins widths approach the size of bands used in their calculations.

\section{Performance with limited wavelength coverage coverage}\label{sec:wave_coverage}
We now evaluate the performance of SQUEzE as a function of wavelength coverage. We explore this by returning to the original data and removing 1/4 and 1/2 of the total coverage. We analyse five cases: red1, red2, blue1, blue2, and mid. Their wavelength coverage is specified in Table~\ref{ta:wave_coverage}.

\begin{table}
    \centering
    \begin{tabular}{cc}
        \toprule
        Sample & Wavelength coverage $\left[\angs\right]$ \\
        \midrule
        orig & 3,600-10,400 \\
        red1 & 5,300-10,400 \\
        red2 & 7,000-10,400 \\
        blue1 & 3,600-8,700 \\
        blue2 & 3,600-7,000 \\
        mid & 5,300-8,700\\
        \bottomrule
    \end{tabular}
    \caption{Wavelength coverage of the original data and the five case studies.}
    \label{ta:wave_coverage}
\end{table}

\begin{figure*}
    \centering
    \includegraphics[width=\textwidth]{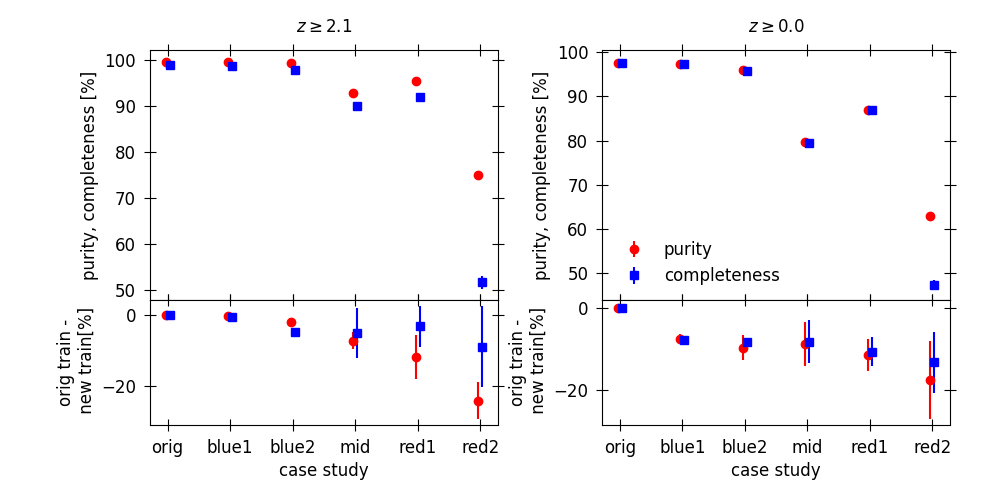}
    \caption{Similar to Figure~\ref{fig:noise_summary} but for the case studies with different wavelength coverage (see Table~\ref{ta:wave_coverage} for details).}
    \label{fig:wave_summary}
\end{figure*}

\begin{figure}
    \centering
    \includegraphics[width=0.47\textwidth]{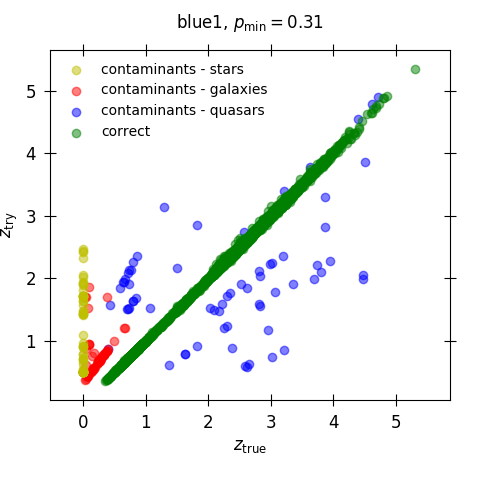}
    \caption{Similar to Figure~\ref{fig:line_confusion_rebin100} but for the case study blue1, where the model has been retrained.}
    \label{fig:line_confusion_blue1}
\end{figure}

\begin{figure}
    \centering
    \includegraphics[width=0.47\textwidth]{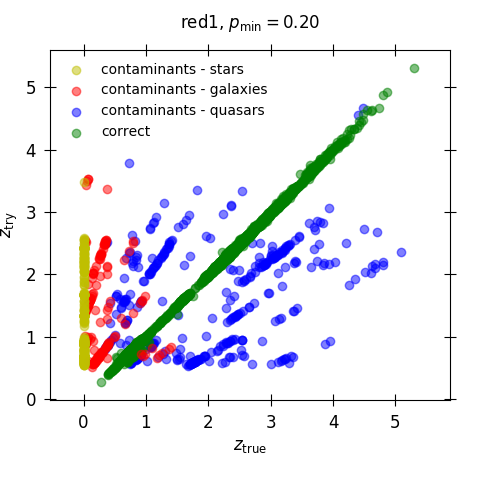}
    \caption{Similar to Figure~\ref{fig:line_confusion_rebin100} but for the case study red1, where the model has been retrained.}
    \label{fig:line_confusion_red1}
\end{figure}

Figure~\ref{fig:wave_summary} summarizes the result of this exercise. SQUEzE performance is strongly reduced for samples mid, red1, and particularly red2. The  samples blue1 and blue2, on the other hand, present a much milder decrease in performance. This suggests that the blue end of the spectra are driving the classification. To better understand this behaviour we analyze the line confusion plots for samples blue1 and red1 (Figures~\ref{fig:line_confusion_blue1} and \ref{fig:line_confusion_red1}, respectively). Note that $p_{\rm min}$ is allowed to vary between wavelength coverage case studies such that each one balances purity and completeness for the sample of all quasars. We see that the decrease on purity and completeness for sample red1 is mostly due to line confusion (note the multiple lines of quasar contaminants with slope far from unity). Losing the blue part of the spectrum results in SQUEzE misidentify quasar emission lines. This is most likely due to the fact that the blue portion of a quasar spectrum (above the Lyman limit in the quasar restframe) is more crowded with major emission lines (see in figure 1 of \citealt{Perez-Rafols+2019}), which typically fall in the blue portion of the observed wavelength range in our sample. The strongest line is of course Lyman-$\alpha$, which motivates our treatment $z>2.1$ quasar targets as a special case.

The lower panels of Figure~\ref{fig:wave_summary} shows the performance decrease if the original training from full wavelength is used instead of retraining on the limited coverage. This shows that SQUEzE performs significantly better upon retraining. This change is easily understood by the behaviour of the code. When SQUEzE attempts to compute the metrics of a line that is outside the covered range, then it assigns it a \texttt{NaN} value. This means that the values of some metrics in the restricted wavelength case will have a \texttt{NaN} value, whereas it will have a non-\texttt{NaN} value in the original training dataset (without the wavelength coverage restriction). 
This substantial difference in the values of metrics with and without retraining has a significant impact. Hence retraining to deal with markedly limited wavelength coverage is recommended.

\section{Performance vs quasar brightness}\label{sec:brightlow}

Thus far we have presented the kind of sample differences that might arise for a survey that observes the same quasar target sample as BOSS but with different observatories, instrumentation and set-ups. Important performance modifications may also arise due to changes in the quasar target sample. We explore this point here by testing the impact of quasar brightness over the dynamic range provided by the BOSS sample.

One of the major assumptions of SQUEzE is that quasars are self-similar, but in reality this is just an approximation: we know that there is a quasar-to-quasar variation. This variation arises from a number of things: presence of broad absorption lines, damped Lyman-$\alpha$ absorbers, difference in continuum slope(s) and differences emission line strengths. All this variation exists in our test samples, but if the survey on which SQUEzE is implemented differs significantly in its realization of these various effects, SQUEzE may not perform as we have described. It is clear that one way future surveys will differ from BOSS is in their luminosity distributions, targeting to a fainter limiting magnitude. Going to fainter magnitudes will most likely change the distribution of quasar properties.
An example of such a change is the well-known Baldwin effect \citep{Baldwin1977}, which states that the relative strength of the emission lines depends on the luminosity.

While we cannot yet test SQUEzE on significant numbers of quasars fainter than the BOSS limit ($r \lesssim 22.2 $), we can test the sensitivity to quasar faintness by taking brightness dependent sample subsets. To this end, we split the quasar target sample into two bright/faint sub-samples of roughly equivalent signal-to-noise by obtaining approximately equal values for the ensemble quantity
\begin{equation}
    \sum\left(\frac{r_{i}}{\delta r_{i}}\right)^{2} ~,
\end{equation}
where $r_{i}$ and $\delta r_{i}$ are the r-band magnitude and its error\footnote{as reported in the superset catalogue file, available at \url{https://data.sdss.org/sas/dr12/boss/qso/DR12Q/Superset_DR12Q.fits}} for the {\it i}th spectrum of that sub-sample.

The boundary which conserves r-band signal to noise between bright and faint samples is $r\sim16$. 
We apply models trained using the bright samples (bright models) to both the bright and the faint validation samples. Similarly we apply models trained using the faint samples (faint models) to both bright and faint validation samples. 

\begin{figure*}
    \centering
    \includegraphics[width=\textwidth]{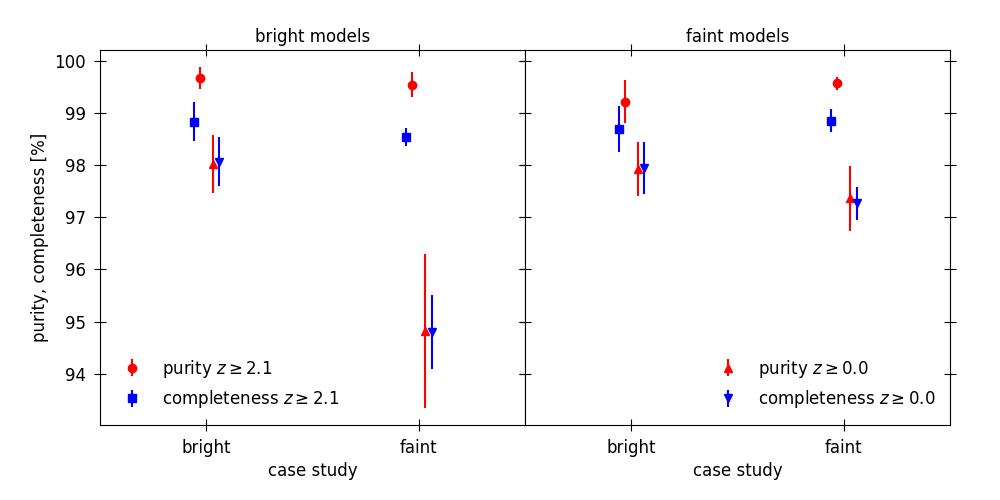}
    \caption{Performance of SQUEzE when applied to samples of different brightness. Left (right) panel shows the result applying the bright (faint) models. See text for details.}
    \label{fig:brightlow}
\end{figure*}

Results of this exercise are shown in Figure~\ref{fig:brightlow}. We observe that SQUEzE performance on $z>2.1$ quasars is not significantly affected by these sample changes. It is particularly striking that SQUEzE performs well on faint $z>2.1$ quasars, including quasars as faint as $r\sim22$, even when it is trained using quasars brighter than $r\sim16$. 

This does not remain the case when all redshifts are included. It would appear that the performance for faint quasars does indeed benefit from retraining on similar quasars but only at the $\sim2-3\%$ level. It appears that this is driven by a dependence on the weaker emission lines with wavelengths long-ward of Lyman $\alpha$.

\section{Performance on upcoming surveys}\label{sec:new_surveys}

The tests we have presented in the previous sections are presented as survey-neutral stress tests that explore how SQUEzE performs given a variety of differences of observing conditions or quasar population. However, they do have direct relevance for real-world expected use-cases for up-coming surveys. In this section we frame our tests with reference to what the mean for the use of SQUEzE on these surveys.

In the near future, the two most relevant surveys are WEAVE-QSO and DESI. In both cases the minimum signal-to-noise to full depth for quasars with $z>2.1$ will be similar to that of BOSS and therefore the results shown in Paper I are already a reasonable guide to quasar catalogue making performance with regards to noise. However, the quasar populations may themselves differ from that of BOSS since these surveys reach a limiting magnitude $\Delta r\sim1$ fainter than BOSS. While we do not assess  the performance for this fainter population we do test the importance of quasar brightness. We find that within the BOSS magnitude distribution, SQUEzE performs satisfactorily upon faint $z \ge 2.1$ quasars even when training using bright $z \ge 2.1$ quasars. When including $z<2.1$ quasars retraining on quasars with representative magnitudes improves performance by $\sim2-3\%$. Both DESI and WEAVE-QSO provide higher spectral resolution than BOSS but our binning tests show that a factor of $\sim 2$ or $\sim 3$ in the resolution respectively will have no impact on SQUEzE performance.

For both DESI and WEAVE-QSO, significant challenges are faced with regards to building the survey. In the case of DESI, only $z > 2.1$ quasars will be observed to full depth in four layers (or passes) over the survey footprint. Quasars must be efficiently identified (and assigned approximate redshifts) based on a single layer (or a quarter of the exposure time) to either determine whether to continue in subsequent observing layers or to refine the redshift with the data as-is. Single layer data is equivalent to our noise4 case (see tests in Section~\ref{sec:noise}). We find that SQUEzE will be able to identify $z \ge 2.1$ quasars on the first pass to high purity but that completeness may be reduced by a few percent (depending on the choice of probability threshold). Including quasars with $z < 2.1$ presents more of a challenge with a larger impact on both purity and completeness (of approximately 7\%) that cannot be recovered trivially by tuning the probability threshold.

In the case of WEAVE-QSO, SQUEzE is also expected to perform well for catalogue production and as a redshift prior for additional redshift refinement step. Applying SQUEzE to the unmodified BOSS sample for $z>2.1$ we see that, assuming there are no dramatic changes to the quasar or contaminant population, we expect almost no losses or impurities for this task. The main challenge for WEAVE-QSO is at the target selection stage, since  $gtr$ 90\% completeness and purity of $z>2.1$ quasars is needed from the quasar targets. The WEAVE-QSO plan is to achieve this goal using data from the survey J-PAS.
J-PAS is a narrow-band photometric survey with 56 narrow-band filters effectively providing spectra of very low resolution. These pseudo-spectra have an effective bin width of $\sim 100\angs$. Our results of the impact of binning show SQUEzE performs well even at the level of $100\angs$ bins (case rebin100). The purity of $z>2.1$  sample seems to be unaffected and the completeness shows only an additional 1\% loss compared to the unmodified sample. The WEAVE-QSO survey is entirely focused on Lyman $\alpha$ forest quasars, but it is noteworthy that the performance of SQUEzE on J-PAS data at all redshifts remains strong.

An additional challenge for WEAVE-QSO is reflected in the tight schedule of J-PAS targeting and WEAVE observing. As a result it is likely that in some instances only some of the filters will be observed in time for propagation to WEAVE-QSO fiber assignment. In this regard SQUEzE performance tests with limited wavelength coverage could be critical. The success of the blue2 test (using only the blue half of the optical range) indicates a potential way forward for the planning for J-PAS and WEAVE-QSO scheduling using SQUEzE. 

\begin{figure*}
    \centering
    \includegraphics[width=\textwidth]{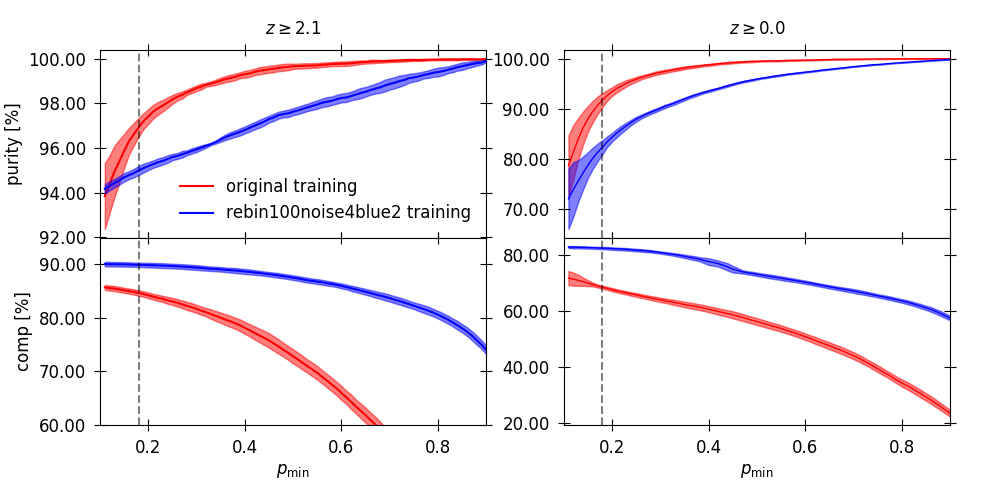}
    \caption{The dependence of the rebin100+noise4+blue2 results using the original or the retrained models for varying probability thresholds. The plot shows the purity (top panels) and the completeness (bottom panels) measured on the test sample with the noise variance increased by a factor of 4, with the pixels rebinned to $100\angs$-wide pixels,  and the wavelength range limited to only the blue half of the spectra (see Table~\ref{ta:wave_coverage}).
    Red and blue lines show the results when the classifier is trained with the original data and with the rebinned data, respectively. Left panels shows the results for $z_{\rm try} \geq 2.1$, whilst right panels show the results at all redshifts. Dashed vertical lines show the value of $p_{\rm min}$ for which purity and completeness are equal for the retrained model and at all redshifts.}
    \label{fig:rebin100noise4blue2}
\end{figure*}

This is a simplification, however, since these are isolated tests. True quasar identification in J-PAS will involve a combination of increased noise (since the limiting magnitude required by WEAVE-QSO will be 1 magnitude fainter than BOSS), $\sim 100$\AA\ binning), and (at times) limited wavelengths coverage. Realistic J-PAS mocks are needed to fully address this challenge. 
In order to take initial steps towards a more realistic test including all these effects simultaneously, we perform the analysis on a modified sample combining the cases rebin100, noise4 and blue2. Our noise4 modification should produce spectra of roughly the expected signal-to-noise of J-PAS pseudo-spectra (Carolina Queiroz, private communication). We stress that here we are  considering the worst-case scenario in which all the data have limited wavelength coverage. The results of this exercise are given in figure \ref{fig:rebin100noise4blue2}. We can see that the performance obtained in this case is lower than the performance obtained in the cases noise4, rebin100, and blue2. Purity and completeness drop below 90\% for all redshift, but we recover a 90\% completeness and 95\% purity for the \lya{} quasars. We note that all these changes combine in a non-trivial way. Interestingly, the purity seems to be higher when the original training is used, but we believe this is a direct consequence of the reduced completeness. For example for $p_{\rm min}=0.3$ and for \lya{} quasars purity is better by $\sim3\%$, but completeness drops by $\sim10\%$, and this is even worse for higher values of $p_{\rm min}=0.3$. A high purity sample can be achieved using the retrained sample with a sufficiently high choice for  $p_{\rm min}$ with less cost in terms of completeness compared to using the original training. The results presented here suggest that SQUEzE will be able to adapt to this challenge, but we leave such detailed tests for future studies.

Other surveys such as Euclid or WEAVE-LOFAR (another survey part of the WEAVE Collaboration) may also benefit from the use of SQUEzE, but their specific requirements have not yet been explored.

\section{Summary}\label{sec:discussion}
In this work we have explored the universality of the performance of SQUEzE (presented in Paper I), a software packaged designed to identify quasar spectra, among a set of contaminating spectra, and estimate their redshifts. We define this universality two ways. Firstly, we confront SQUEzE with the more straightforward challenge of training on and processing data with different spectroscopic properties (compared to BOSS data in Paper I) and explore how these changes affect performance. Secondly, we assess the universality of SQUEzE {\it combined with} the training established in Paper I by confronting BOSS-trained SQUEzE with different data. This latter test allows us to assess the sensitivity to the realism of the training set, and even exploring whether SQUEzE needs retraining on the different data at all.

We have addressed the effect of changing the signal-to-noise, the spectral binning (and resolution), the wavelength coverage and the brightness of the sample. After all these tests we are confident that the performance of SQUEzE is largely survey-independent. However, for optimal performance we do recommend altering BOSS spectra to resemble the characteristics of the survey to be analysed and retraining the model on the modified data. If more precise and reliable tests are required, a more realistic contamination sample matched to the details of the given survey must be explored (such a sample could be obtained, for example, in the survey validation phase). We do not recommend retraining using synthetic spectra, unless full modelling of the contaminants is also taken into account.

All the tests performed here assume that the visual inspection is always correct, but this is not necessarily the case. We will explore this in more detail in a subsequent paper of the series. However, we remark that the tests are fair since what we analyse is the difference in the results on the original and modified spectra, and whatever miss-classifications are present in truth table for the original samples are also present in the modified samples.

Finally, we make initial tests of SQUEzE performance on upcoming surveys. We conclude that SQUEzE will perform satisfactorily on spectra from DESI (including single layer spectra from the first pass), and WEAVE-QSO. We also explore the case of J-PAS pseudo-spectra, which will be used for the target selection of WEAVE-QSO, reaching the same conclusion. Other surveys such as Euclid or WEAVE-LOFAR may also benefit from the use of SQUEzE, but their specific requirements have not yet been explored.

\section*{Acknowledgments}
We thank Carolina Queiroz for very helpful discussions on the expected quality of J-PAS data. 

This work was partly  supported by the A*MIDEX project (ANR-11-IDEX-0001-02) funded by the ``Investissements d'Avenir'' French Government program, managed by the French National Research Agency (ANR), and by ANR under contract ANR-14-ACHN-0021.

\section*{Data availability}
The data underlying this article is available in SDSS repository at \url{https://data.sdss.org/sas/dr12/boss/spectro/redux/v5_7_0/spectra/lite/}.
The code and the standard training originally presented in Paper I is publicly available at \url{https://github.com/iprafols/SQUEzE}.

\bibliographystyle{apj}
\bibliography{iprafols}

\end{document}